\documentclass[amsmath,amssymb,12pt]{revtex4-1}
\usepackage[utf8]{inputenc} 
\usepackage{graphicx}
\usepackage{float}
\usepackage{natbib}
\usepackage{wrapfig}
\usepackage{xcolor}
\usepackage[english]{babel}
\begin{document}

	\title{Matrix coupling and generalized frustration in Kuramoto oscillators}

	\author{Guilhermo L. Buzanello} 
	
	\author{Ana Elisa D.  Barioni} 
	
	\author{Marcus A. M. de Aguiar }
	
	\affiliation{Instituto de F\'isica `Gleb Wataghin', Universidade Estadual de Campinas, Unicamp 13083-970, Campinas, SP, Brazil}
	
	\begin{abstract}
		
		The Kuramoto model describes the synchronization of coupled oscillators that have different natural frequencies. Among the many generalizations of the original model, Kuramoto and Sakaguchi (KS) proposed a {\it frustrated} version that resulted in dynamic behavior of the order parameter, even when the average natural frequency of the oscillators is zero. Here we consider a generalization of the frustrated KS model that exhibits new transitions to synchronization. The model is identical in form to the original Kuramoto model, but written in terms of unit vectors. Replacing the coupling constant by a coupling matrix breaks the rotational symmetry and forces the order parameter to point in the direction of the eigenvector with highest eigenvalue, when the eigenvalues are real. For complex eigenvalues the module of order parameter oscillates while it rotates around the unit circle, creating active states. We derive the complete phase diagram for Lorentzian distribution of frequencies using the Ott-Antonsen ansatz. We also show that changing the average value of the natural frequencies leads to further phase transitions where the module of the order parameter goes from oscillatory to static.
		
	\end{abstract}
	
	\maketitle

	\section{Introduction}
	
	In 1975 Kuramoto  proposed a simple model of $N$ coupled oscillators that could be solved analytically in the limit where $N$ goes to infinity \cite{Kuramoto1975,Kuramoto1984}. The oscillators are described by their phases $\theta_i$ 
	and are coupled according to the equations 
	\begin{equation}
		\dot{\theta}_i = \omega_i + \frac{K}{N} \sum_{j=1}^N \sin{(\theta_j-\theta_i)}
		\label{kuramoto}
	\end{equation}
	where  $\omega_i$ are their natural frequencies, selected from a symmetric distribution $g(\omega)$, $K$ is the coupling strength and $i = 1, ..., N$.  The complex order parameter 
	\begin{equation}
		z = p e^{i \psi} \equiv \frac{1}{N} \sum_{i=1}^N e^{i\theta_i}
		\label{paraord}
	\end{equation}
	measures the degree of phase synchronization of the particles: disordered motion implies $p \approx 0$ and coherent motion $p \approx 1$. The phase of $z$ has been largely ignored, as it converges to a constant that depends on the initial conditions. 
	Random initial conditions lead, therefore, to random values of $\psi$ for each simulation. 
	
	Kuramoto showed that the onset of synchronization could be described, in equilibrium, as a continuous phase transition, where $p$ remains very small for $0 < K < K_c$ and increases as $p = \sqrt{1-K_c/K}$ for $ K>K_c$. Since then, a large number of modifications and generalizations of his model have been proposed, including different types of coupling functions \cite{hong2011kuramoto,yeung1999time,breakspear2010generative}, introduction of networks of connections (so that not all oscillators are connected to each other) \cite{Rodrigues2016,Joyce2019},
	different distributions of the oscillator's natural frequencies (including frequencies proportional to the number of connections, leading to explosive synchronization) \cite{Gomez-Gardenes2011,Ji2013}, inertial terms \cite{Acebron2005,dorfler2011critical,olmi2014hysteretic}, external periodic driving forces
	\cite{Childs2008,moreira2019global,moreira2019modular} and coupling with particle swarms \cite{o2017oscillators,o2022collective}.
	
	More recently, interest has been shifted to understand oscillations in larger dimensions. Chandra et al \cite{chandra2019continuous} have shown that Kuramoto oscillators could also be described by unit vectors $\vec{\sigma_i} =
	(\cos{\theta_i},\sin{\theta_i})$ that rotate on the unit circle. It is easy to show that if  $\theta_i$ satisfies
	Kuramoto's equation (\ref{kuramoto}) then 
	\begin{equation}
		\frac{d \vec{\sigma_i}}{d t} = \mathbf{W}_i \vec{\sigma_i} + \frac{K}{N} \sum_j [\vec{\sigma_j} - (\vec{\sigma_i}\cdot \vec{\sigma_j}) \vec{\sigma_i}]
		\label{eq3}
	\end{equation}
	where  $\mathbf{W}_i$ is an anti-symmetric matrix containing the natural frequency $\omega_i$:
	\begin{equation}
		\mathbf{W}_i = \left( 
		\begin{array}{cc}
			0 & -\omega_i \\
			\omega_i & 0
		\end{array}
		\right).
		\label{wmat}
	\end{equation}
	The complex order parameter  $z$, Eq.(\ref{paraord}), is replaced by the vector
	\begin{equation}
		\vec{p} = \frac{1}{N}\sum_i \vec{\sigma_i} = (p\cos\psi,p\sin\psi)
		\label{vecpar}
	\end{equation}
	describing the center of mass of the system. Eq.(\ref{eq3}) can be naturally extended to higher dimensions by simply considering unit vectors $\vec\sigma_i$ in D-dimensions rotating on the surface of the corresponding (D-1)-sphere. It has been shown, in particular, that the system exhibits discontinuous  phase transitions in odd dimensions and continuous transitions in even dimensions
	\cite{chandra2019continuous}. Also, complexity reduction similar to that proposed by Ott and Antonsen \cite{Ott2008} can be applied in any dimension \cite{chandra2019complexity,barioni2021complexity,barioni2021ott}. 
	
	In this paper we explore the original two-dimensional case (Section II) generalizing the coupling constant $K$ to a $2 \times 2$ matrix \textbf{K} that can be separated into a rotation plus a symmetric matrix. The corresponding Kuramoto equations are then written as a generalization of the Kuramoto-Sakaguchi model. This allows us to construct a complete phase diagram (Section III) displaying three different regions: no synchrony, phase tuned states and active states. In addition, varying the average frequency of the oscillators we find conditions for the stability of these state. Finally, we compare our analytical model to the numerical simulations (Section IV).

	\section{Matrix coupling}
	
	Kuramoto's model in vector form can be naturally extended if the coupling constant $K$ in Eq.(\ref{eq3}) is replaced  by a $2 \times 2$ matrices ${\mathbf K}_{ij} $  with elements that might depend also on the angles $\phi_i$ and $\phi_j$. The equations now read
	\begin{equation}
		\frac{d \vec{\sigma_i}}{d t} = \mathbf{W}_i \vec{\sigma_i} + \frac{1}{N} \sum_j [{\mathbf K}_{ij} \vec{\sigma_j} - (\vec{\sigma_i}\cdot {\mathbf K}_{ij} \vec{\sigma_j}) \vec{\sigma_i}]
		\label{kuragen}
	\end{equation}
	and can be interpreted as a generalized frustrated model, as $\mathbf{K}_{ij}$ rotates $\vec\sigma_j$ hindering its alignment with $\vec\sigma_i$ and inhibiting synchronization.
	
	Defining the auxiliary vectors $\vec{q}_{i} = \sum_j {\mathbf K}_{ij} \, \vec{\sigma}_j$ we can rewrite theses equations as
	\begin{equation}
		\frac{d \vec{\sigma_i}}{d t} = \mathbf{W}_i \vec{\sigma_i} +  [ \vec{q}_{i} - (\vec{\sigma_i}\cdot \vec{q}_{i}) \vec{\sigma_i}].
		\label{kuramotogenk}
	\end{equation}
	Norm conservation, $|\vec{\sigma_i}|=1$, is guaranteed for any set of regular matrices ${\mathbf K}_{ij}$, as can be seen by taking the scalar product of Eqs.(\ref{kuragen}) or (\ref{kuramotogenk}) with $\vec{\sigma_i}$. In this paper we shall only consider the case where ${\mathbf K}_{ij}={\mathbf K}= constant$ and $\vec{q}_{i} = \vec{q} = \mathbf{K} \vec{p}$. Writing 
	\begin{equation}
		\mathbf{K} = \left( 
		\begin{array}{cc}
			a & b \\
			c & d
		\end{array}
		\right)
		\label{kmat}
	\end{equation}
	Eq.(\ref{kuragen}) can be written back in terms of the phases $\theta_i$ as
	\begin{equation}
		\dot{\theta}_i = \omega_i + \frac{1}{2N} \sum_{j=1}^N \left[ (a+d) \sin{\theta_{ji}^-}  + (d-a) \sin{\theta_{ji}^+} + (c-b) \cos{\theta_{ji}^-} + (b+c) \cos{\theta_{ji}^+} \right]
		\label{kuragentheta}
	\end{equation}
	where ${\theta_{ji}^-} = \theta_j - \theta_i$ and ${\theta_{ji}^+} = \theta_j + \theta_i$. From this representation we immediately recognize two special cases: (i) $a=d$, $b=c=0$, corresponding to the usual Kuramoto model and, (ii) $a=d=K \cos \alpha$, $b=-c=K\sin \alpha$ (a rotation matrix). In this case the terms inside the brackets in Eq.(\ref{kuragentheta}) simplify to $\sin(\theta_j - \theta_i - \alpha)$, corresponding to the Kuramoto-Sakaguchi model \cite{sakaguchi1986soluble,yue2020model}. This motivates us to re-parametrize $\mathbf{K}$ as
	\begin{equation}
		\mathbf{K} = K
		\left( 
		\begin{array}{cc}
			\cos\alpha & \sin\alpha \\
			-\sin\alpha & \cos\alpha
		\end{array}
		\right) + 
		J \left( 
		\begin{array}{cc}
			-\cos\beta & \sin\beta \\
			\sin\beta & \cos\beta
		\end{array}
		\right) \equiv \mathbf{K}_R  + \mathbf{K}_S  
		\label{kgen}
	\end{equation}
	where $\mathbf{K}_R$ is a rotation and $\mathbf{K}_S$ a symmetric matrix. In these variables Eq.(\ref{kuragentheta}) reads
	\begin{equation}
		\dot{\theta}_i = \omega_i + \frac{1}{N} \sum_{j=1}^N \left[ K \sin(\theta_j - \theta_i - \alpha) + J \sin(\theta_j + \theta_i + \beta) \right].
		\label{kuragenjk}
	\end{equation}
	and the eigenvalues of $\mathbf{K}$ are given by
	\begin{equation}
		\lambda_\pm = K\cos\alpha \pm \sqrt{J^2-K^2\sin^2\alpha}
		\label{eigen}    
	\end{equation}
	and are independent of $\beta$.

	\subsection{Continuity equation}
	
	In the limit of infinitely many oscillators we define $f(\omega,\theta,t)$ as the density of oscillators with natural frequency $\omega$ 
	at position $\theta$ in time $t$. It satisfies the continuity equation
	\begin{equation}
		\frac{\partial f}{\partial t} +  \frac{\partial (f v_\theta)}{\partial \theta}   = 0
		\label{cont1}
	\end{equation}
	with velocity field
	\begin{equation}
		\vec{v} =  \vec{\omega} \times \hat{r} +  \vec{q} - (\hat{r} \cdot \vec{q}) \hat{r}= (\omega + q_\theta) \hat{\theta}  \equiv v_\theta \hat{\theta}
	\end{equation}
	where $\vec{\omega}$ points perpendicular to the plane of rotation.  Eq.(\ref{vecpar}) for the order parameter becomes 
	\begin{equation}
		\vec{p}(t) = \int \hat{r}(\theta) f(\omega,\theta,t) d \theta  \, d \omega.
		\label{param}
	\end{equation}
	
	\subsection{OA ansatz for density function}
	
	The density of oscillators is a periodic function of $\theta$ and, therefore, can be expanded in Fourier series. Ott \&
	Antonsen \cite{Ott2008} showed that if Fourier coefficients are chosen as $\rho^{|m|} e^{-im\phi}$ the solution is
	self-consistent, in the sense that it preserves this form at all times, remaining in this restricted subset of density
	functions. The ansatz, parametrized by $\rho(\omega,t)$ and $\phi(\omega,t)$ is: 
	\begin{eqnarray}
		f(\omega,\theta,t) = \frac{g(\omega)}{2\pi} \sum_{m=-\infty}^{\infty}  \rho^{|m|}  e^{im(\theta-\phi)} = 
		\frac{g(\omega)}{2\pi} \frac{(1-\rho^2)}{1+\rho^2-2 \rho \cos (\theta - \phi)}.
		\label{f1}
	\end{eqnarray}
	It is convenient to define the vector $\vec{\rho} = \rho (\cos \phi, \sin \phi)$ so that $\cos (\theta-\phi) = \hat{r} \cdot \hat{\rho}$.  Substituting Eq.(\ref{f1}) into (\ref{param}) we find:
	\begin{equation}
		\vec{p}(t) = \int \vec{\rho}(\omega,t)  g(\omega) \, d \omega.
		\label{paramg}
	\end{equation}
	Also, substituting Eq.(\ref{f1}) into Eq.(\ref{cont1}) we obtain \cite{barioni2021complexity}
	\begin{eqnarray}
		\dot{\vec{\rho}}  =  \vec{\omega} \times \vec{\rho} + \frac{1}{2}(1+\rho^2) {\mathbf K} \vec{p} -  ({\mathbf K}  \vec{p} \cdot \vec{\rho})  \vec{\rho}.
		\label{eqmf}
	\end{eqnarray}

	\subsection{Complex variables}
	
	The Ott-Antonsen ansatz is particularly useful for Lorentzian distributions $g(\omega)$. In order to derive the equation satisfied by the order parameter in this case we define the complex variables
	\begin{equation}
		z = pe^{i\psi}, \qquad w = \rho e^{i\phi}, \qquad u = q e^{i\xi}.
	\end{equation}
	Eq.(\ref{eqmf}) then becomes
	\begin{equation}
		\dot{w} = i \omega w + \frac{u}{2} - \frac{u^*}{2} w^2.
		\label{zeq}
	\end{equation}
	where
	\begin{equation}
		u = Kze^{-i\alpha} - J z^* e^{-i\beta}
		\label{ueq}    
	\end{equation}
	is obtained from the definition $\vec{q} = \mathbf{K} \vec{p}$. Substituting (\ref{ueq}) into (\ref{zeq}) we obtain
	\begin{equation}
		\dot{w} = i \omega w + \frac{1}{2}\left(Kze^{-i \alpha} - J z^* e^{-i \beta} \right) - \frac{w^2}{2} \left(Kz^*e^{i \alpha} - J z e^{i \beta} \right).
		\label{zeq2}
	\end{equation}
	Equation (\ref{paramg}), on the other hand becomes
	\begin{equation}
		z = \int w(\omega) g(\omega) d \omega.
		\label{orderpar}
	\end{equation}
	We note that if we define $F = -J z^* e^{-i \beta}$, Eq.(\ref{zeq2}) can be written as
	\begin{equation}
		\dot{w} = i \omega w + \frac{1}{2}\left(Kze^{-i \alpha} +F  \right) - \frac{w^2}{2} \left(Kze^{-i \alpha} + F \right)^*
		\label{zeq3}
	\end{equation}
	which is identical to the forced Kuramoto model discussed in \cite{Childs2008}, with the difference that now the `force' $F$ is generated by the system itself, instead of being applied externally. 
	
	\section{Dynamics and phase diagram}
	
	For the Lorentzian distribution 
	\begin{equation}
		g(\omega) = \frac{1}{\pi} \frac{\Delta}{(\omega-\omega_0)^2 + \Delta^2}
	\end{equation}
	Eq.(\ref{orderpar}) can be integrated in the complex plane using a close contour from $\omega=-R$ to $\omega=R$ and back to $-R$ through the upper half circle $\omega=R e^{i\phi}$, enclosing the pole at $\omega= \omega_0 + i\Delta$. Taking $R\rightarrow \infty$ we obtain $z = w(\omega_0 + i \Delta)$ \cite{Ott2008}. Calculating Eq.(\ref{zeq2}) at $\omega_0 + i \Delta$ we can replace $w$ by $z$ to get 
	\begin{equation}
		\dot{z} = i (\omega_0  + i \Delta)z + \frac{1}{2}\left(Kze^{-i \alpha} - J z^* e^{-i \beta} \right) - \frac{z^2}{2} \left(Kz^*e^{i \alpha} - J z e^{i \beta} \right).
		\label{zeq4}
	\end{equation}
	Separating real and imaginary parts we obtain
	\begin{equation}
		\dot{p} = -\Delta p + \frac{p}{2} \left( 1 - p^2\right) \left[ K\cos\alpha - J \cos(2\psi+\beta) \right] 
		\label{peq}    
	\end{equation}
	and
	\begin{equation}
		\dot{\psi} = \omega_0 - \frac{1}{2} \left( 1 + p^2\right) \left[ K\sin\alpha - J \sin(2\psi+\beta) \right].
		\label{psieq}    
	\end{equation}
	
	\subsection{Equilibrium solutions}
	
	Strict equilibrium solutions of Eqs.(\ref{peq}) and (\ref{psieq}) are only possible if $J=\alpha=\omega_0=0$. In this case $\psi$ is constant and Eq.(\ref{peq}) is autonomous. However, a less stringent definition of equilibrium can be obtained by setting only $J=0$. In this case Eq.(\ref{peq}) has two stationary solutions, $p=0$ and $p=\sqrt{1-K_c/K}$, where $K_c \equiv 2\Delta/\cos\alpha$. The latter bifurcates at $K=K_c$ and corresponds to the synchronized state of the Kuramoto-Sakaguchi model. At this solution $\vec{p}$ rotates with constant angular velocity $\dot\psi = \omega_0 -K \sin\alpha + \Delta \tan \alpha$. It is then possible to change to a frame of reference that rotates with $\psi$, making the system stationary. This is possible because $\mathbf{K}$ is itself a rotation matrix and commutes with the operation that changes reference frames.
	
	\subsection{Phase tuned states - real eigenvalues}
	
	If $|J| > |K| \sin \alpha$ the eigenvalues of $\mathbf{K}$ are real (see Eq.(\ref{eigen})). Let $\vec{v} = (\cos\gamma,\sin\gamma)$ be the eigenvector corresponding to the largest eigenvalue of $\mathbf{K}$. From $\mathbf{K} \vec{v} = \lambda_+ \vec{v}$ it follows that $K\sin\alpha-J\sin(2\gamma+\beta) = 0$. This implies that when $\vec{p}$ is an eigenvector of $\mathbf{K}$ ($\psi=\gamma$) we find $\dot\psi = \omega_0$. We call this region {\it phase tuned}, as the phase $\psi$ can be tuned with the choice of $\mathbf{K}$. Eq. (\ref{kuramotogenk}) then seems to recover the Kuramoto model with $\lambda_+$ playing the role of the scalar coupling constant $K$, as $\vec{q} = \mathbf{K} \vec{p} = \lambda_+ \vec{p}$. This, however, is not so. To see why, let us consider the trivial solution $p=0$ of Eq.(\ref{peq}). Linear stability analysis leads to the equation
	\begin{equation}
		\delta \dot{p} = -\delta p \left( \Delta  -   \frac{K}{2} \cos\alpha + \frac{J}{2} \cos(2\psi+\beta) \right)
		\label{pstab}    
	\end{equation}
	which becomes unstable if
	\begin{equation}
		K \cos\alpha - J \cos(2\psi+\beta) \geq 2\Delta
	\end{equation}
	Using $K\sin\alpha=J\sin(2\psi+\beta)$ (as $\vec{p}$ is eigenvector of $\mathbf{K}$) we can rewrite this condition as 
	\begin{equation}
		K\cos\alpha + \sqrt{J^2-K^2\sin^2\alpha} \geq 2\Delta
		\label{eigen2}    
	\end{equation}
	which corresponds to set $\lambda_+ \geq \Delta$.
	
	The solution that branches off from $p=0$ at $\lambda_+ = \Delta$ is 
	\begin{equation}
		p = \sqrt{1-\frac{2\Delta}{K\cos\alpha - J \cos(2\psi - \beta)}}
	\end{equation}
	but this is only stationary if $\psi$ is constant ($\omega_0=0$) or $J=0$. For $\omega_0 \neq 0$, there is no nontrivial equilibrium solution. The trick of changing to a new frame of reference rotating with $\omega_0$ does not work, as $\mathbf{K}$ does not commute with rotations if $J\neq 0$. Such operation would change $\mathbf{K}$ to $\mathbf{\tilde{K}} = \mathbf{R}^T \mathbf{K} \mathbf{R}$ which would be itself time dependent. For $\omega_0 \neq 0$ the order parameter branches off to an active state where it rotates while its module oscillates, as if an external force were acting on the system (see next subsection). We recall the interpretation of Eq.(\ref{zeq3}) where $J$ indeed acts as an external drive to the system. For $\omega_0 = 0$ the order parameter remains stationary and pointing in the direction of the eigenvector of $\mathbf{K}$, breaking the rotational symmetry present in the original Kuramoto model. Some of these results can be demonstrated for any symmetric distribution of frequencies $g(\omega)$, not just the Lorentzian (see Appendix A).

	\subsection{Active states - complex eigenvalues}
	
	\begin{figure*}
		\centering 
		\includegraphics[scale=0.5]{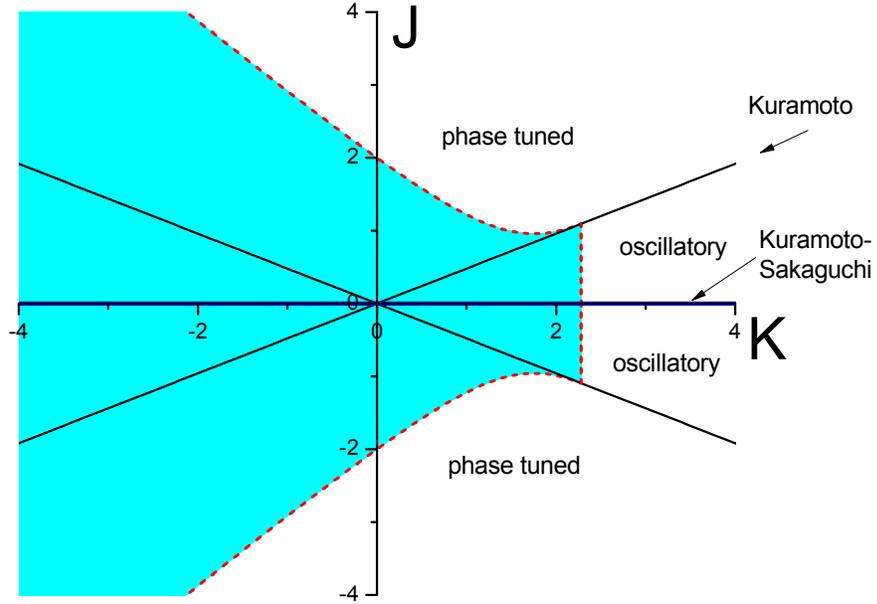} 
		\caption{Phase diagram for $\alpha=0.5$ and $\Delta=1$. The red line separates synchronized motion (to the right) from disordered motion (shaded bluled area to the left). Diagonal lines with slope equal to $\sin\alpha$ separate regions with complex eigenvalues (enclosing the line $J=0$) from real eigenvalues (enclosing $K=0$). In the phase tuned region, the phase of the order parameter points in the direction of the eigenvector of $\mathbf{K}$ with the largest eigenvalue if $\omega_0=0$. In the oscillatory region $\vec{p}$ rotates and its module oscillates even if $\omega_0=0$ (active states). Exactly at the diagonals the eigenvalues of $\mathbf{K}$ are degenerated and the system behaves like the original Kuramoto model. On the K-axis ($J=0$) we recover the Kuramoto-Sakaguchi model.} 
		\label{fig1}
	\end{figure*}
	
	\begin{figure*}
		\centering 
		\includegraphics[scale=0.5]{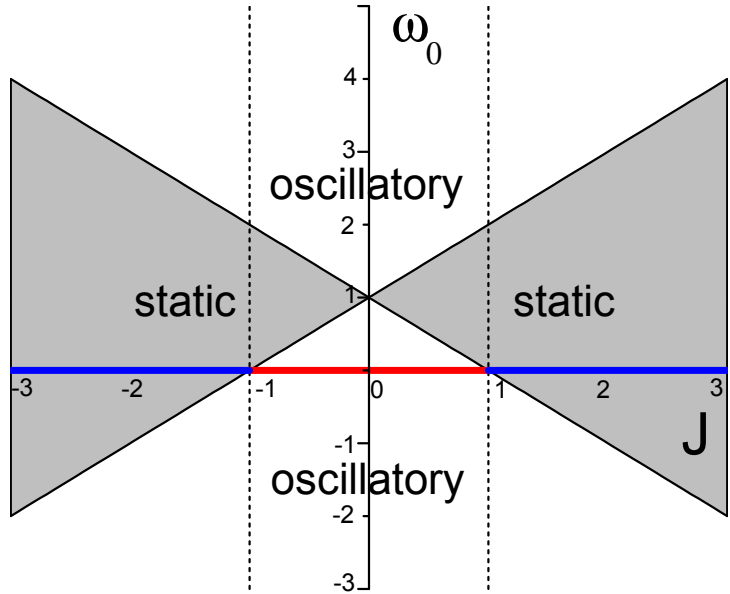} 
		\caption{Phase diagram in the plane $J-\omega_0$ for fixed $K$ and $\alpha$. The x-axis is plotted in units of $K\sin\alpha$ and the y-axis in units of $K\sin\alpha/\zeta$. For $\omega_0=0$ and $|J|< K\sin\alpha$ (red line) the system is in the oscillatory region. As $\omega_0$ is increased and go inside the gray cone generated by the black lines (Eq.(\ref{zeta})) $p$ stops oscillating (Fig. \ref{fig4}(a)). For $|J| > K\sin\alpha$ and $\omega_0=0$ (blue lines) the system is phase tuned region with $\vec{p}$ pointing in the direction of the eigenvector of $\mathbf{K}$ with largest eigenvalue. As $\omega_0$ increases or decreases and go outside the gray cone, $p$ starts oscillating (Fig. \ref{fig4}(b)).} 
		\label{fig2}
	\end{figure*}
	
	If $|J| < |K| \sin \alpha$ the eigenvalues of $\mathbf{K}$ are complex and the relation $K\sin\alpha-J\sin(2\gamma+\beta) = 0$ does not hold, as the eigenvectors are also complex. Still, the trivial solution $p=0$ satisfies Eq.(\ref{peq}). For $\psi$ we find 
	\begin{equation}
		\dot{\psi} = \omega_0 - \frac{1}{2} \left[ K\sin\alpha - J \sin(2\psi+\beta) \right]
		\label{eqpsi0}    
	\end{equation}
	whose solution is
	\begin{equation}
		(2\omega_0-K\sin\alpha) \tan(\psi+\beta/2) = -J + 2\Omega \tan[\Omega(t-t_0)]
		\label{psisol}
	\end{equation}
	where $\Omega=\sqrt{(2\omega_0-K\sin\alpha)^2-J^2}/2$. For $\omega_0=0$ this corresponds to the imaginary part of $\lambda$, generalizing the Kuramoto-Sakaguchi dynamics. For $\omega_0 \neq 0$, oscillations in $\psi$ (and therefore in the modulus of $\vec{p}$) require $\Omega$ to be real, otherwise $\psi$ goes to a constant that can be computed from Eq.(\ref{psisol}) replacing $\Omega \tan[\Omega(t-t_0)]$ by $|\Omega| \tanh[|\Omega|(t-t_0)]$. Therefore, oscillatory solutions can be `stabilized' by choosing $\omega_0$ in the window 
	\begin{equation}
		\frac{K\sin\alpha -|J|}{\zeta} < \omega_0 < \frac{K\sin\alpha+|J|}{\zeta}.
		\label{zeta}
	\end{equation}
	where $\zeta=2$. The boundaries of this region correspond to phase transitions from oscillatory to constant behavior of $\vec{p}$. Notice that these equations for $\psi$ were derived at the trivial solution $p=0$. However, we can assume that they are also valid for $p\neq 0$, as this would only change factor $\zeta=2$ in Eq.(\ref{eqpsi0}) to $2/(p+1)$, which can be approximated by its time-averaged value $\zeta = 2/(\langle p \rangle +1)$. A similar effect occurs in the phase-tuned region, where $\omega_0$ can induce oscillations in $p$. 
	
	The solution of Eq.(\ref{pstab}) for $J \neq 0$ is
	\begin{equation}
		\delta p(t) = \delta p(0) \exp{ \left\{ \left(\frac{K}{2}\cos\alpha - \Delta \right)t + \frac{J}{2} \int_0^t \cos(2\psi(t') + \beta) dt' \right\}}.
	\end{equation}
	Because $\cos(2\psi(t) +\beta)$ oscillates around zero, the integral in the exponent remains finite and the condition for $p=0$ to be unstable is
	\begin{equation}
		K \cos\alpha  \geq 2\Delta
	\end{equation}
	which is independent of $J$. Figure \ref{fig1} shows the complete phase diagram in the $J-K$ plane for $\omega_0=0$ displaying the regions where the three different types of solutions occur. A diagram illustrating the transitions as a function of $\omega_0$ is shown in Fig.\ref{fig2}.
	
	\begin{figure*}
		\centering 
		\includegraphics[scale=0.28]{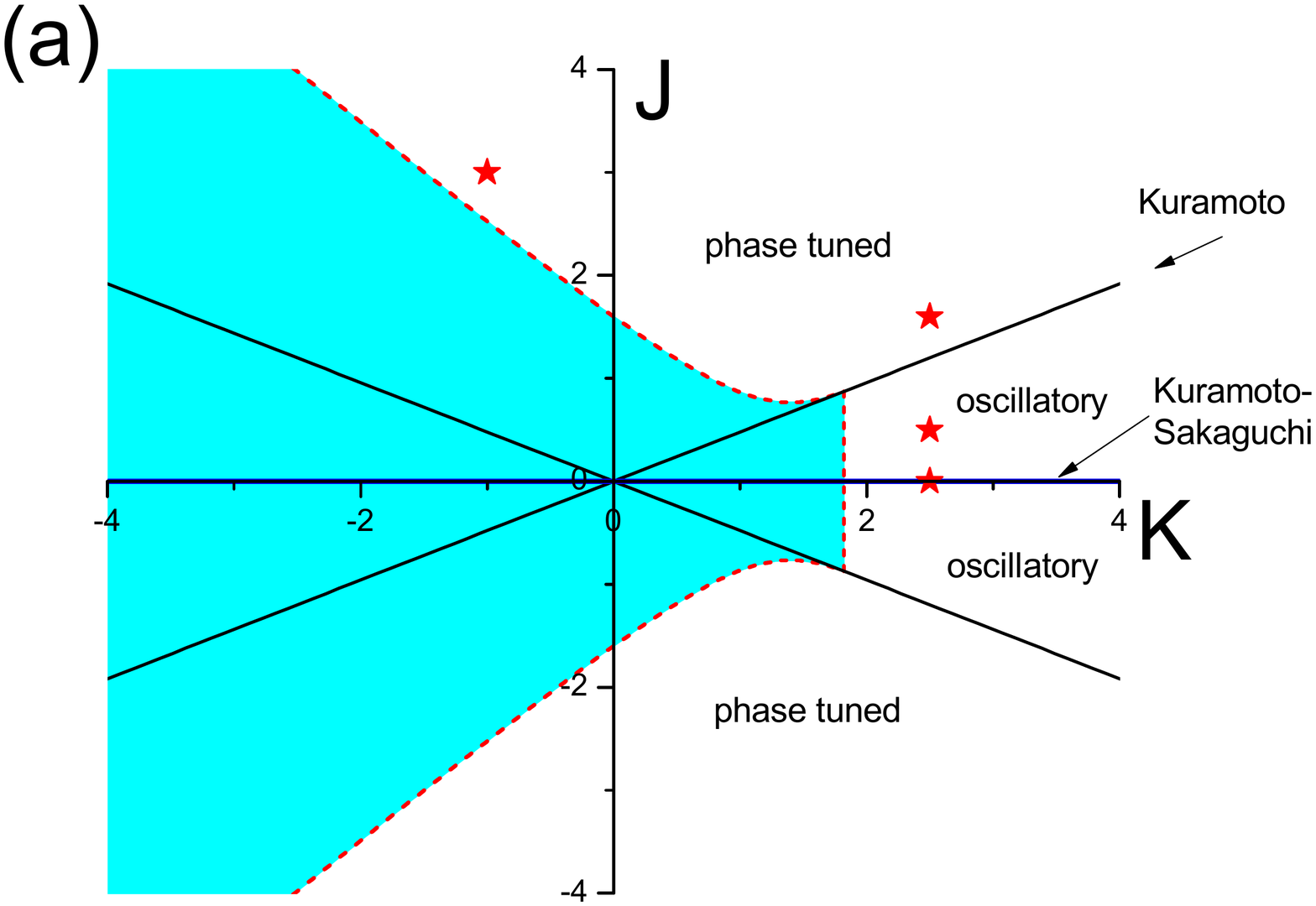} 
		\includegraphics[scale=0.28]{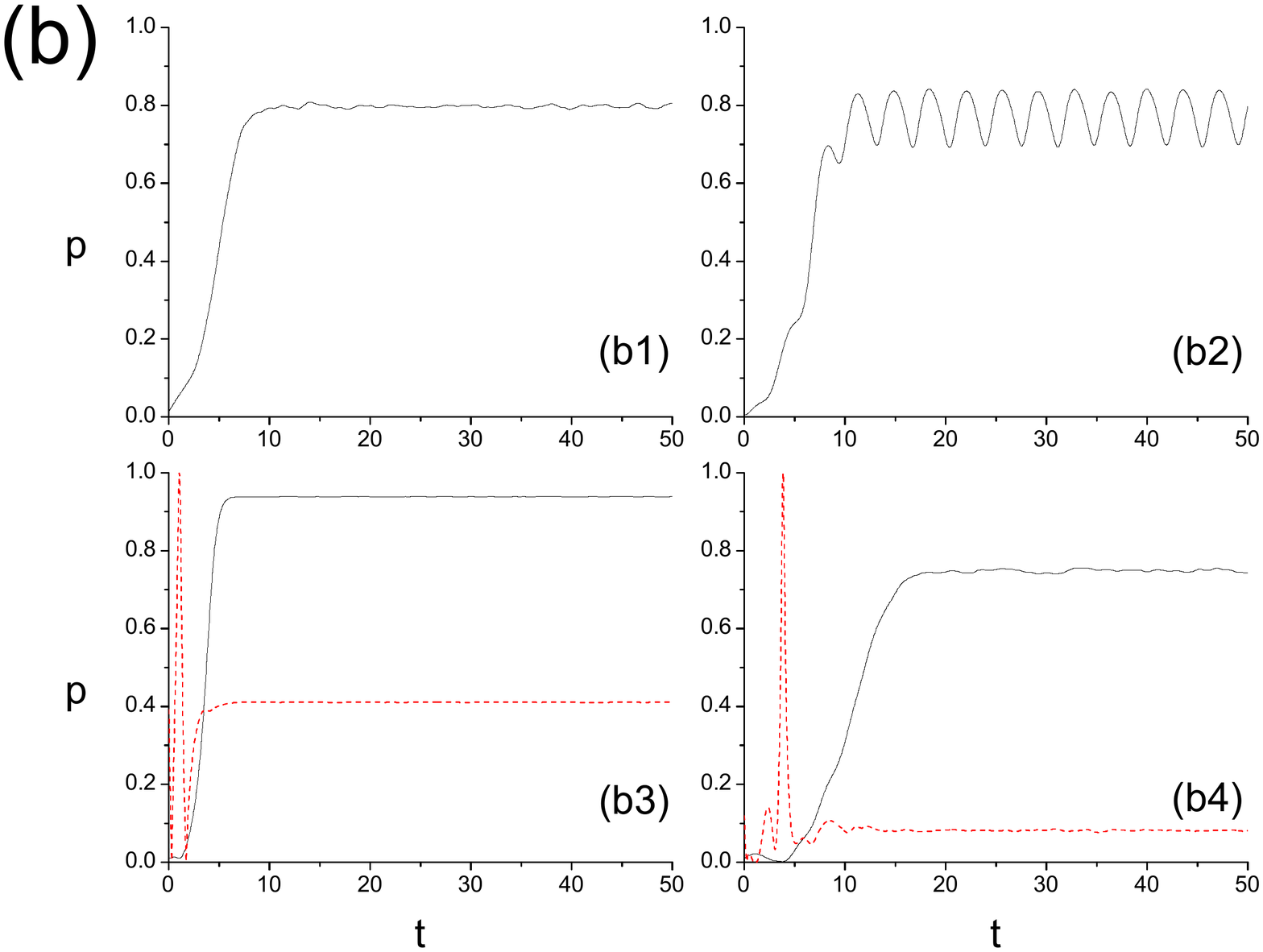} 
		\caption{(a) Phase diagram for $\alpha=0.5$ and Gaussian distribution of frequencies with $\omega_0=0$. Stars show the values of $K$ and $J$ displayed on the plots in panel (b): (b1) $K=2.5; J=0$  (Kuramoto-Sakaguichi); (b2) $K=2.5; J=0.5$ (active oscillatory state); (b3) $K=2.5; J=1.6$ (phase tuned state); (b4) $K=-1; J=3$ (phase tuned state with negative $K$. The red lines in panels (b3) and (b4) show $\cos\psi$ and correspond to the direction of the eigenvector of $\mathbf{K}$ with largest eigenvalue.} 
		\label{fig3}
	\end{figure*}
	
	\section{Simulations}
	
	The theoretical diagrams described in Figs. \ref{fig1} and \ref{fig2} can be confirmed by numerical simulations. In this section we show results for $N=10000$ oscillators, $\alpha=0.5$ and Gaussian distribution of natural frequencies, as the system converges faster in this case than with Lorentzian distributions. Fig. \ref{fig3}(a) shows the phase diagram in the $J-K$ plane adapted to Gaussian distributions. Red stars show parameter values selected for the simulations shown in Fig. \ref{fig3}(b). In the active region $p$ oscillates (panel b2) unless $J=0$ (panel b1). In the phase tuned region (panels b3 and b4) $p$ converges to a constant value, and so does $\psi$ (red lines), as $\vec{p}$ becomes parallel to the eigenvector of $\mathbf{K}$ with largest eigenvalue.
	
\begin{figure*}
		\centering 
		\includegraphics[scale=0.28]{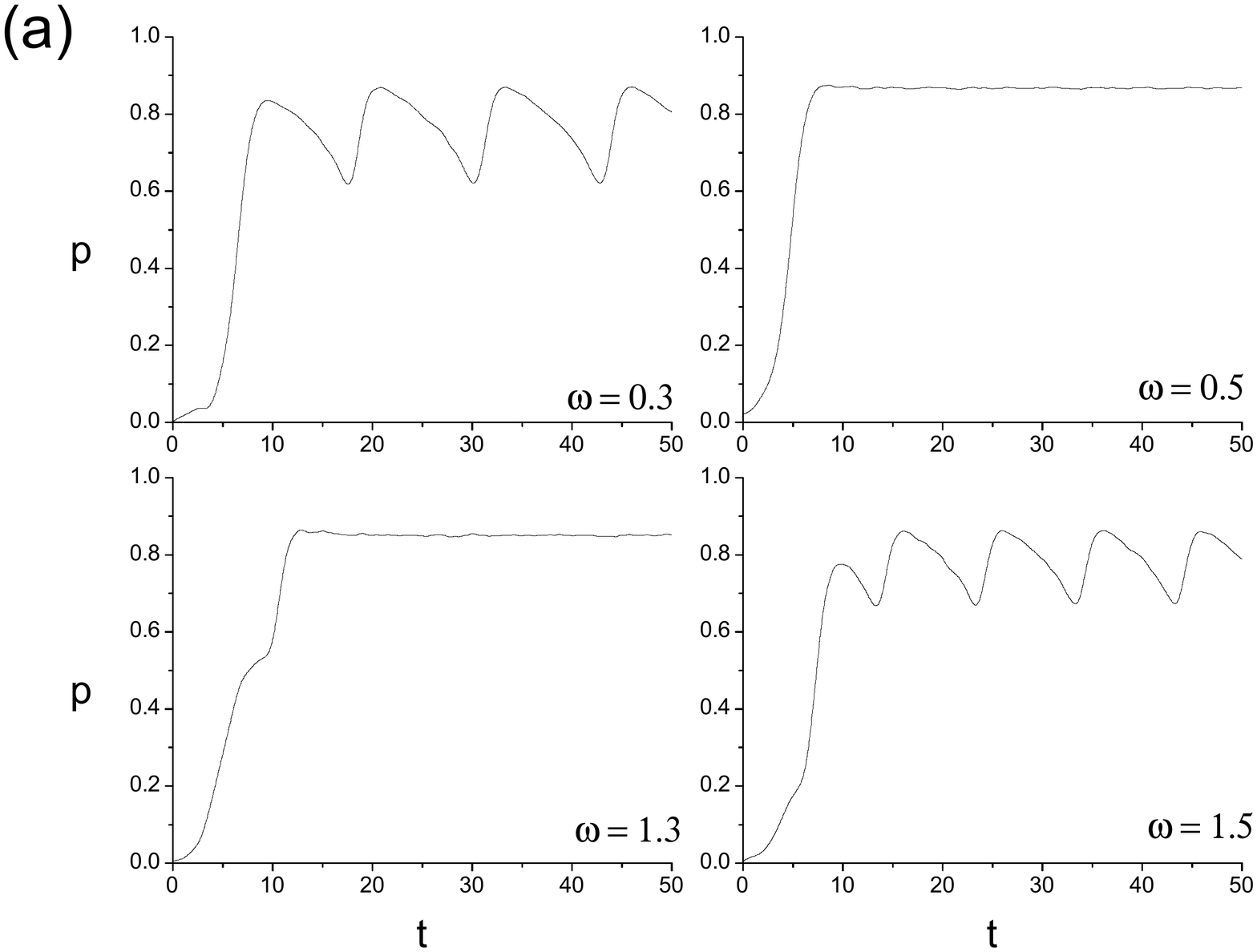} 
		\includegraphics[scale=0.28]{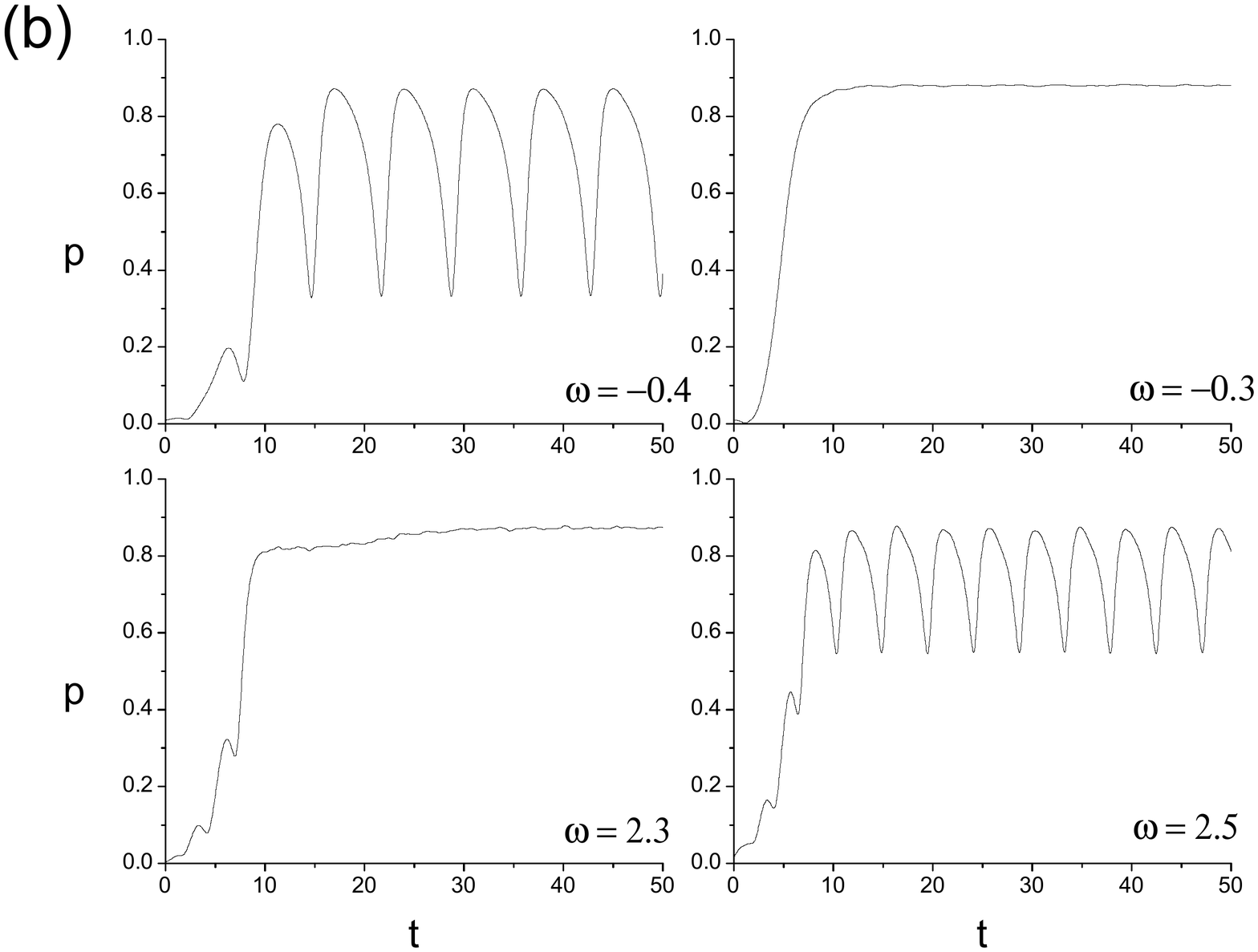} 
		\caption{(a) Transition from oscillatory to static and back to oscillatory behavior as $\omega_0$ is varied in the active oscillatory region for $K=2.5$ and $J=0.5$. Values of $\omega_0$ are indicated in the panels. (b) Similar transitions for the phase tuned region for $K=2.5$ and $J=1.6$ . } 
		\label{fig4}
	\end{figure*}

	Fig. \ref{fig4} shows the transitions induced by $\omega_0$. Panel (a) displays results for the active region with $K=2.5$ and $J=0.5$. For $\omega_0=0.3$ the oscillations in $p$ persist. For $0.6 < \omega_0 < 1.4$ the oscillations are dumped and $p$ converges to a constant. However, for $\omega_0 > 1.4$ the system oscillates again (see Fig. \ref{fig2}). Panel (b) shows similar results for the phase tuned region with $K=2.5$ and $J=1.6$. In this case $p$ remains constant for $-0.35 < \omega_0 < 2.4$. Outside this range $p$ acquires oscillations induced by $\omega_0$.
	
	\section{Conclusions}
	
	We have extended the Kuramoto model by writing the dynamic equations in vector form \cite{chandra2019continuous} and promoting the scalar coupling constant $K$ to a $2\times 2$ matrix $\mathbf{K}$. Splitting the matrix into a rotation $\mathbf{K}_R$  plus a symmetric matrix $\mathbf{K}_S$ we were able to write down the equations governing the evolution of the order parameter in the limit of infinite oscillators and Lorentzian distributions of natural frequencies. We showed that $\mathbf{K}_R$  gives rise to the Kuramoto-Sakaguchi model whereas the symmetric part $\mathbf{K}_S$ acts as internal driving force that breaks the rotational symmetry. 
	
	We constructed the complete phase diagram for the case where the average frequency of the oscillators, $\omega_0$, is zero. Solutions are divided into {\it disordered} (no synchrony),  {\it phase tuned} (where module and phase of the order parameter converges to specific values) -- and {\it active states} (where module and direction of the order parameter oscillate in time). The original Kuramoto model and the frustrated model of Kuramoto and Sakaguchi are special cases of our model. Finally, we showed that non-zero values of $\omega_0$ induce further transitions in the behavior of the order parameter, creating a cone in the $J$-$\omega_0$ plane (Fig. \ref{fig2}) inside of which the solutions are static (constant $p$) and outside are oscillatory.
	
	This novel generalization of the Kuramoto model opens a broad range of modifications to be explored and studied. Examples include oscillators of higher dimensions, networks replacing the all-to-all connections of the model and applications to swarmalators \cite{o2017oscillators}. We hope the contributions discussed in this paper can lead to new interpretations and to deeper comprehension of the Kuramoto model and its applications.

	\begin{acknowledgments}
		This work was partly supported by FAPESP, grants 2019/24068-0 (AEDB), 2021/10709-3 (GLB), 2016/01343‐7 (MAMA, ICTP‐SAIFR) and CNPq, grant 301082/2019‐7 (MAMA). 
	\end{acknowledgments}
	
	\begin{appendix}
		
		\section{General distribution of natural frequencies}
		
		In this appendix we show that the critical line separating asynchronous and synchronous states can be obtained for general distributions of natural frequencies $g(\omega)$ using more qualitative arguments. For this we go back to Eq.(\ref{eqmf}) and take the scalar product with $\hat{\rho}$ to get
		\begin{eqnarray}
			\dot{\rho} &=& \frac{1}{2} (1-\rho^2)  ({\mathbf K} \vec{p} \cdot \hat{\rho}).
			\label{eqmff}
		\end{eqnarray}
		Equilibrium requires that either $\rho=1$, and we only need to find the direction $\hat{\rho}$, or $\vec{\rho}$ is perpendicular to $\mathbf{K} \vec{p}$. For $\rho=1$ we obtain
		\begin{equation}
			0 = \vec{\omega} \times \hat{\rho} + {\mathbf K} \vec{p} - ({\mathbf K} \vec{p} \cdot \hat{\rho}) \hat{\rho}.
		\end{equation}
		Taking the cross product with $\hat{\rho}$ we get
		$\vec{\omega}  = - \hat{\rho} \times {\mathbf K} \vec{p}$ and $\omega = q \sin(\phi -\xi)$, which requires $|\omega| < q$. 
		
		The tree important vectors are: the order parameter $\vec{p} = (p\cos\psi,p\sin\psi)$, the ansatz vector $\vec{\rho} = (\rho\cos\phi,\rho\sin\phi)$ and the auxiliary vector $\vec{q} = {\mathbf K} \vec{p} = (q\cos\xi,q\sin\xi)$. Using $d \omega = q \cos(\phi-\xi) d \phi$ we obtain, for the x-component of Eq.(\ref{paramg})
		\begin{equation}
			p \cos\psi = \int_{-q}^{+q} \cos \phi \, g(\omega) d \omega = \int_{\xi -\pi/2}^{\xi+\pi/2} q \cos \phi \cos(\phi -\xi) g(\omega(\phi)) d \phi
		\end{equation}
		as the solutions for $|\omega| > q$ (corresponding to ${\mathbf K} \vec{p} \cdot \hat{\rho} = 0$) do not contribute to $\vec{p}$. The trivial solution corresponds to $p=0$, as this implies $q=0$. The critical condition for synchronization corresponds to the non-trivial solution calculated at $p=0$. Evaluating $g(\omega)$ at $q=0$ and doing the integral over $\phi$ we get
		\begin{equation}
			p \cos\psi = ( g(0)  \pi/2) q \cos\xi
			\label{cospsi}
		\end{equation}
		and, similarly,
		\begin{equation}
			p \sin\psi = ( g(0)  \pi/2) q \sin\xi.
			\label{sinpsi}
		\end{equation}
		Dividing one equation by the other we find that $ \tan \xi = \tan \psi$, implying that $\vec{q}$ is parallel to $\vec{p}$, or ${\mathbf K} \vec{p} = \lambda \vec{p}$. The direction of the order parameter is therefore fixed by coupling matrix if its eigenvalues are real. Moreover, replacing $q$ by $\lambda p$ in Eqs. (\ref{cospsi}) and (\ref{sinpsi}) we find the critical condition for synchronization as 
		\begin{equation}
			\lambda_c = 2/(\pi g(0))
			\label{lambdac}
		\end{equation}
		which is identical to the condition found for the Lorenztian distribution, where $g(0)=(\Delta \pi)^{-1}$. 
		
		%
		%
		%
		%
		
	\end{appendix}
	
	\clearpage 
	\newpage

\begin{thebibliography}{25}
		\providecommand{\natexlab}[1]{#1}
		\providecommand{\url}[1]{\texttt{#1}}
		\expandafter\ifx\csname urlstyle\endcsname\relax
		\providecommand{\doi}[1]{doi: #1}\else
		\providecommand{\doi}{doi: \begingroup \urlstyle{rm}\Url}\fi
		
		\bibitem[Kuramoto(1975)]{Kuramoto1975}
		Yoshiki Kuramoto.
		\newblock {Self-entrainment of a population of coupled non-linear oscillators}.
		\newblock In \emph{International Symposium on Mathematical Problems in
			Theoretical Physics}, pages 420--422. Springer-Verlag, Berlin/Heidelberg,
		1975.
		\newblock \doi{10.1007/BFb0013365}.
		\newblock URL \url{http://www.springerlink.com/index/10.1007/BFb0013365}.
		
		\bibitem[Kuramoto(1984)]{Kuramoto1984}
		Yoshiki Kuramoto.
		\newblock {Chemical Waves}.
		\newblock In \emph{Chemical Oscillations, Waves, and Turbulence}, pages
		89--110. Springer Berlin Heidelberg, 1984.
		\newblock \doi{10.1007/978-3-642-69689-3_6}.
		\newblock URL
		\url{http://www.springerlink.com/index/10.1007/978-3-642-69689-3{\_}6}.
		
		\bibitem[Hong and Strogatz(2011)]{hong2011kuramoto}
		Hyunsuk Hong and Steven~H Strogatz.
		\newblock Kuramoto model of coupled oscillators with positive and negative
		coupling parameters: an example of conformist and contrarian oscillators.
		\newblock \emph{Physical Review Letters}, 106\penalty0 (5):\penalty0 054102,
		2011.
		
		\bibitem[Yeung and Strogatz(1999)]{yeung1999time}
		MK~Stephen Yeung and Steven~H Strogatz.
		\newblock Time delay in the kuramoto model of coupled oscillators.
		\newblock \emph{Physical Review Letters}, 82\penalty0 (3):\penalty0 648, 1999.
		
		\bibitem[Breakspear et~al.(2010)Breakspear, Heitmann, and
		Daffertshofer]{breakspear2010generative}
		Michael Breakspear, Stewart Heitmann, and Andreas Daffertshofer.
		\newblock Generative models of cortical oscillations: neurobiological
		implications of the kuramoto model.
		\newblock \emph{Frontiers in human neuroscience}, 4:\penalty0 190, 2010.
		
		\bibitem[Rodrigues et~al.(2016)Rodrigues, Peron, Ji, and Kurths]{Rodrigues2016}
		Francisco~A. Rodrigues, Thomas K D~M Peron, Peng Ji, and J??rgen Kurths.
		\newblock {The Kuramoto model in complex networks}.
		\newblock \emph{Physics Reports}, 610:\penalty0 1--98, 2016.
		\newblock ISSN 03701573.
		\newblock \doi{10.1016/j.physrep.2015.10.008}.
		\newblock URL \url{http://dx.doi.org/10.1016/j.physrep.2015.10.008}.
		
		\bibitem[Climaco and Saa(2019)]{Joyce2019}
		Joyce~S. Climaco and Alberto Saa.
		\newblock Optimal global synchronization of partially forced kuramoto
		oscillators.
		\newblock \emph{Chaos: An Interdisciplinary Journal of Nonlinear Science},
		29\penalty0 (7):\penalty0 073115, 2019.
		\newblock \doi{10.1063/1.5097847}.
		\newblock URL \url{https://doi.org/10.1063/1.5097847}.
		
		\bibitem[Gomez-Gardenes et~al.(2011)Gomez-Gardenes, Gomez, Arenas, and
		Moreno]{Gomez-Gardenes2011}
		Jesus Gomez-Gardenes, Sergio Gomez, Alex Arenas, and Yamir Moreno.
		\newblock {Explosive synchronization transitions in scale-free networks}.
		\newblock \emph{Physical Review Letters}, 106\penalty0 (12):\penalty0 1--4,
		2011.
		\newblock ISSN 00319007.
		\newblock \doi{10.1103/PhysRevLett.106.128701}.
		
		\bibitem[Ji et~al.(2013)Ji, Peron, Menck, Rodrigues, and Kurths]{Ji2013}
		Peng Ji, Thomas K~Dm Peron, Peter~J. Menck, Francisco~A. Rodrigues, and J??rgen
		Kurths.
		\newblock {Cluster explosive synchronization in complex networks}.
		\newblock \emph{Physical Review Letters}, 110\penalty0 (21):\penalty0 1--5,
		2013.
		\newblock ISSN 00319007.
		\newblock \doi{10.1103/PhysRevLett.110.218701}.
		
		\bibitem[Acebr{\'{o}}n et~al.(2005)Acebr{\'{o}}n, Bonilla, Vicente, Ritort, and
		Spigler]{Acebron2005}
		Juan~A. Acebr{\'{o}}n, L.~L. Bonilla, Conrad J~P{\'{e}}rez Vicente, F{\'{e}}lix
		Ritort, and Renato Spigler.
		\newblock {The Kuramoto model: A simple paradigm for synchronization
			phenomena}.
		\newblock \emph{Reviews of Modern Physics}, 77\penalty0 (1):\penalty0 137--185,
		2005.
		\newblock ISSN 00346861.
		\newblock \doi{10.1103/RevModPhys.77.137}.
		
		\bibitem[D{\"o}rfler and Bullo(2011)]{dorfler2011critical}
		Florian D{\"o}rfler and Francesco Bullo.
		\newblock On the critical coupling for kuramoto oscillators.
		\newblock \emph{SIAM Journal on Applied Dynamical Systems}, 10\penalty0
		(3):\penalty0 1070--1099, 2011.
		
		\bibitem[Olmi et~al.(2014)Olmi, Navas, Boccaletti, and
		Torcini]{olmi2014hysteretic}
		Simona Olmi, Adrian Navas, Stefano Boccaletti, and Alessandro Torcini.
		\newblock Hysteretic transitions in the kuramoto model with inertia.
		\newblock \emph{Physical Review E}, 90\penalty0 (4):\penalty0 042905, 2014.
		
		\bibitem[Childs and Strogatz(2008)]{Childs2008}
		Lauren~M. Childs and Steven~H. Strogatz.
		\newblock {Stability diagram for the forced Kuramoto model}.
		\newblock \emph{Chaos}, 18\penalty0 (4):\penalty0 1--9, 2008.
		\newblock ISSN 10541500.
		\newblock \doi{10.1063/1.3049136}.
		
		\bibitem[Moreira and de~Aguiar(2019{\natexlab{a}})]{moreira2019global}
		Carolina~A Moreira and Marcus~AM de~Aguiar.
		\newblock Global synchronization of partially forced kuramoto oscillators on
		networks.
		\newblock \emph{Physica A: Statistical Mechanics and its Applications},
		514:\penalty0 487--496, 2019{\natexlab{a}}.
		
		\bibitem[Moreira and de~Aguiar(2019{\natexlab{b}})]{moreira2019modular}
		Carolina~A Moreira and Marcus~AM de~Aguiar.
		\newblock Modular structure in c. elegans neural network and its response to
		external localized stimuli.
		\newblock \emph{Physica A: Statistical Mechanics and its Applications},
		533:\penalty0 122051, 2019{\natexlab{b}}.
		
		\bibitem[O’Keeffe et~al.(2017)O’Keeffe, Hong, and
		Strogatz]{o2017oscillators}
		Kevin~P O’Keeffe, Hyunsuk Hong, and Steven~H Strogatz.
		\newblock Oscillators that sync and swarm.
		\newblock \emph{Nature communications}, 8\penalty0 (1):\penalty0 1--13, 2017.
		
		\bibitem[O'Keeffe et~al.(2022)O'Keeffe, Ceron, and Petersen]{o2022collective}
		Kevin O'Keeffe, Steven Ceron, and Kirstin Petersen.
		\newblock Collective behavior of swarmalators on a ring.
		\newblock \emph{Physical Review E}, 105\penalty0 (1):\penalty0 014211, 2022.
		
		\bibitem[Chandra et~al.(2019{\natexlab{a}})Chandra, Girvan, and
		Ott]{chandra2019continuous}
		Sarthak Chandra, Michelle Girvan, and Edward Ott.
		\newblock Continuous versus discontinuous transitions in the d-dimensional
		generalized kuramoto model: Odd d is different.
		\newblock \emph{Physical Review X}, 9\penalty0 (1):\penalty0 011002,
		2019{\natexlab{a}}.
		
		\bibitem[Ott and Antonsen(2008)]{Ott2008}
		Edward Ott and Thomas~M. Antonsen.
		\newblock {Low dimensional behavior of large systems of globally coupled
			oscillators}.
		\newblock \emph{Chaos}, 18\penalty0 (3):\penalty0 1--6, 2008.
		\newblock ISSN 10541500.
		\newblock \doi{10.1063/1.2930766}.
		
		\bibitem[Chandra et~al.(2019{\natexlab{b}})Chandra, Girvan, and
		Ott]{chandra2019complexity}
		Sarthak Chandra, Michelle Girvan, and Edward Ott.
		\newblock Complexity reduction ansatz for systems of interacting orientable
		agents: Beyond the kuramoto model.
		\newblock \emph{Chaos: An Interdisciplinary Journal of Nonlinear Science},
		29\penalty0 (5):\penalty0 053107, 2019{\natexlab{b}}.
		
		\bibitem[Barioni and de~Aguiar(2021{\natexlab{a}})]{barioni2021complexity}
		Ana Elisa~D Barioni and Marcus~AM de~Aguiar.
		\newblock Complexity reduction in the 3d kuramoto model.
		\newblock \emph{Chaos, Solitons \& Fractals}, 149:\penalty0 111090,
		2021{\natexlab{a}}.
		
		\bibitem[Barioni and de~Aguiar(2021{\natexlab{b}})]{barioni2021ott}
		Ana Elisa~D Barioni and Marcus~AM de~Aguiar.
		\newblock Ott--antonsen ansatz for the d-dimensional kuramoto model: A
		constructive approach.
		\newblock \emph{Chaos: An Interdisciplinary Journal of Nonlinear Science},
		31\penalty0 (11):\penalty0 113141, 2021{\natexlab{b}}.
		
		\bibitem[Sakaguchi and Kuramoto(1986)]{sakaguchi1986soluble}
		Hidetsugu Sakaguchi and Yoshiki Kuramoto.
		\newblock A soluble active rotater model showing phase transitions via mutual
		entertainment.
		\newblock \emph{Progress of Theoretical Physics}, 76\penalty0 (3):\penalty0
		576--581, 1986.
		
		\bibitem[Yue et~al.(2020)Yue, Smith, and Gottwald]{yue2020model}
		Wenqi Yue, Lachlan~D Smith, and Georg~A Gottwald.
		\newblock Model reduction for the kuramoto-sakaguchi model: The importance of
		nonentrained rogue oscillators.
		\newblock \emph{Physical Review E}, 101\penalty0 (6):\penalty0 062213, 2020.
		
		
	\end{thebibliography}

\end{document}